\documentstyle[epsf]{article}

\title{
Numerical analysis of
the bond-random antiferromagnetic $S=1$ Heisenberg chain
}
\author{Yoshihiro Nishiyama \\
{\it Graduate School of Science, Osaka University}\\
{\it Machikaneyama, Toyonaka 560, Japan}}
\date{}

\begin{document}
\begin{large}

\maketitle

\section*{Abstract}
Ground state of 
the
bond-random antiferromagnetic
$S=1$ Heisenberg chain with the biquadratic interaction
$- \beta \sum_i ( {\bf S}_i \cdot {\bf S}_{i+1} )^2$
is investigated by means of the exact-diagonalization method and
the finite-size-scaling analysis.
It is shown that the Haldane phase $\beta \approx 0$ 
persists against the randomness;
namely, no randomness-driven phase transition
is observed until at a point of extremely-broad-bond distribution.
We found that in the Haldane phase,
the magnetic correlation length is kept hardly changed.
These results are contrastive to those of an analytic theory
which predicts a second-order
phase transition between the Haldane and the random-singlet
phases at a certain critical randomness.

\section{Introduction}
\label{section1}

The ground-state phase
transition of the disordered quantum system has attracted much
attention recently \cite{Sachdev94}.
Because the transition is driven solely by the quantum fluctuation,
the transition lies in different
universality from the finite-temperature transitions.
The randomness contributes significantly to the anisotropy of the
real-space and the imaginary-time directions.
Thereby, the randomness would also promote new types of
phase transition.

For this purpose, 
the ground state of the 
random quantum spin system has been studied considerably,
see the article \cite{Sachdev94,Rieger96} for a review.
In one dimension, in particular, a real-space-decimation
procedure \cite{Ma79,Bhatt82} yields various definite predictions
\cite{Fisher92,Fisher94,Fisher95}.
For the Heisenberg model in the presence of
the bond randomness \cite{Fisher94,Nagaosa87,Doty92}, for example,
the theory tells that infinitesimal randomness
drives the ground state to
the random-singlet phase,
where the averaged ground-state correlation decays obeying the power law,
\begin{equation}
\left[ 
\langle S^\alpha_0 S^\alpha_r \rangle
\right]_{\rm av}
\sim
1/r^2\ .
\end{equation}
Note that the correlation decays faster than that of the pure chain $\sim1/r$.
Numerical simulations followed 
\cite{Haas93,Runge94,Young96} so as to confirm the analytic predictions.
It would be noteworthy that the $S=1/2$ spin system is viewed as a
hard-core boson system \cite{Matsuda70}.
The random spin system is, thus, equivalent to the 
strongly repulsive boson system embedded in a certain randomness,
which is of the current interest \cite{Fisher89}.

In these studies, the magnitude of the constituent 
spin is set to be one half.
In one dimension, however, there exist some other sorts of magnets
such as the {\it integer}-spin Heisenberg chain and the $S=1/2$ Heisenberg
ladder model.
The ground states of these magnets are different from 
the ground state of the $S=1/2$
Heisenberg chain \cite{Haldane83,Dagotto92}:
In the former case,
a finite magnetic excitation gap opens above
the ground state, and the magnetic correlation decays exponentially
in the ground state.
In the present paper, we investigate the
bond-random antiferromagnetic $S=1$ spin chain.
The Hamiltonian is given by
\begin{equation}
{\cal H}=\sum_{i=1}^{L} J_i \left\{
{\bf S}_i \cdot {\bf S}_{i+1} -\beta
({\bf S}_i \cdot {\bf S}_{i+1})^2
\right\}.
\label{Hamiltonian}
\end{equation}
The operators ${\bf S}_i=(S^x_i,S^y_i,S^z_i)$ denote the $S=1$ spin operators
acting on the site $i$, and satisfy the periodic boundary condition, 
${\bf S}_{L+1}={\bf S}_1$.
We used the following bond distribution,
see Fig. \ref{bond_distribution},
\begin{figure}[htbp]
\begin{center}\leavevmode
\epsfxsize=10cm
\epsfbox{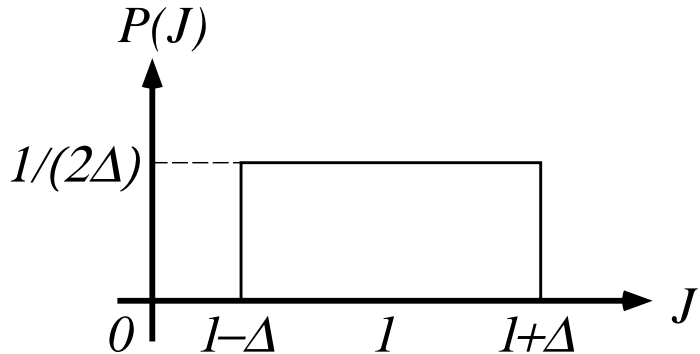}
\end{center}
\caption{
The probability distribution of the random bond $J$.
The randomness ranges as $0 \le \Delta \le 1$.
At $\Delta=1$, the distribution becomes extremely broad;
namely, infinitesimally weak bond appears.
}
\label{bond_distribution}
\end{figure}
\begin{equation}
P_{\Delta} (J)=\frac{1}{2\Delta}
\left(
\Theta(J-1+\Delta)-\Theta(J-1-\Delta)
\right)\ .
\end{equation}
(The function $\Theta(x)$ denotes the step function.)
Note that only the antiferromagnetic bond appears.

Because of the presence of the excitation gap,
the system (\ref{Hamiltonian})
lies beyond the scope of the previous studies of the
$S=1/2$ magnets
explained above.
It is quite suggestive that
the real-space decimation theory
becomes inapplicable for the cases other than $S=1/2$.
In order to adapt the decimation procedure even for the case $S=1$,
several authors proposed 
schemes to map the random $S=1$ chain to an effective $S=1/2$ model
\cite{Boechat96,Hyman97}.
Their theory and the consequences are reviewed afterwards.

The effect of the {\it non-magnetic} impurity doping upon the spin-gap
ground state has been considered
rather throughly \cite{Fukuyama96,Martins96,Mikeska97}:
The non-magnetic impurity causes notable influences
for the Heisenberg ladder.
Infinitesimal doping distracts the spin gap, and gives rise to a
magnetically (quasi-)ordered ground state with gapless excitation.
The effect of the bond randomness, which is of the present interest,
would be more subtle, and remained unsolved.

For the pure system $\{ J_i=1 \}$,
the ground-state phase diagram of the Hamiltonian (\ref{Hamiltonian})
is known, see the paper \cite{Fath95} for details.
We explain the region $\beta\sim0$ relevant to the present study.
In the region $-1 < \beta < 1$, the system is in the Haldane phase,
which is mentioned above; namely,
a finite excitation gap opens, and the magnetic
correlation decays exponentially in the ground state.
% At the point $\beta=-1/3$, the ground state is expressed explicitly.
% For this state, the above properties are confirmed rigorously.
At the critical point $\beta=1$, the Hamiltonian is integrable.
It is predicted in a field-theoretical manner \cite{Affleck85}
that the criticality is of
the central charge $c=3/2$, and
the biquadratic interaction generates the excitation gap in the form,
\begin{equation}
\Delta E \propto |\beta - 1|^\nu\ ,
\end{equation}
where $\nu=1$.
Numerical simulations have tried to confirm the above scenario
\cite{Blote86,Oitmaa86,Saitoh87,Fath95}.
A strong finite-size corrections, however,
arise so as to prevent definite conclusions.
Typical simulation claims that a transition locates at $\beta\approx0.5$,
and gapless phase extends over $0.5<\beta$ \cite{Oitmaa86,Saitoh87}.
The inconsistency might be originated in the 
long correlation length $\xi$ in the
Barber-Batchelor phase $1<\beta$;
$\xi\approx21$ at $\beta\to\infty$ \cite{Barber89}.
The length apparently 
exceeds the limit tractable by means of the exact-diagonalization method.

The paper is organized as follows.
In the next section, we explain the previous analytic theory
employing the real-space-decimation method \cite{Boechat96,Hyman97}.
The theory predicts that
a phase transition separates the Haldane and
the random-singlet phases at a certain critical randomness.
In the section \ref{section3}, we present our result
by means of the exact diagonalization method.
We show that the finite-size scaling assuming the theoretical prediction
fails; namely,
the present numerical estimate of the transition point and the
critical exponent contradicts the prediction.
The Haldane phase persists against the randomness until the extremely-%
broad-bond distribution $\Delta=1$.
The results of the correlation length and the spin stiffness
confirm the present conclusion.
In the last section, we summarize our results,
and discuss a possible scenario interpreting our numerical results.

\section{Review of the predictions with the 
real-space decimation analysis}
\label{section2}

In this section, we review the predictions of the analytic theory
\cite{Boechat96,Hyman97}.
Some are contrastive to our numerical
results shown later.
As was introduced in the section \ref{section1},
the real-space-decimation method describes the one-dimensional
random magnets quite successfully.
The method, however, fails in the cases other than $S=1/2$.
Hyman and Yang thus proposed the mapping with which the $S=1$ random chain
(\ref{Hamiltonian})
is transformed to an effective
random $S=1/2$ chain.
The resultant $S=1/2$ chain constitutes the alternating
two types of random bonds, $\{J_{2i}\}$ and $\{J_{2i-1}\}$.
The even bonds are either 
ferromagnetic or antiferromagnetic, whereas the odd
bonds are always antiferromagnetic.

We sketch how the effective $S=1/2$ random chain is 
derived.
Suppose a sector within the $S=1$ chain in which sector
the deviation of the randomness is incidentally small.
They assumed that such sector
could be replaced with that without any randomness.
As a consequence, the random $S=1$ chain consists of
finite uniform $S=1$ sectors coupling antiferromagnetically.
On the other hand, it was reported that
at an end of the uniform $S=1$ open chain,
the $S=1/2$ magnetization appears spontaneously
with a certain localization length
\cite{Miyashita93}.
Both the end magnetizations
couple antiferromagnetically (ferromagnetically),
if the length of the open chain is even (odd).
Thereby, the finite uniform $S=1$ sector is replaced
with
the two $S=1/2$ spins coupling either antiferromagnetically or
ferromagnetically.

The real-space decimation procedure is employed afterwards.
They found a fixed point at a critical randomness $\Delta_{\rm c}$,
which separates the Haldane phase and the random-singlet phase
\cite{Hyman97}.
(Refer to the section \ref{section1} for these phases.)
Furthermore, the theory concludes that the correlation length 
diverges as
\begin{equation}
\xi \sim 1/|\Delta - \Delta_{\rm c}|^\nu
\label{correlation_exponent}
\end{equation}
with $\nu\approx2.3$.

The decimation procedure was simulated numerically as well \cite{Monthus97}.
The simulation shows that an intermediate phase appears
between the Haldane and the random-singlet phases.

\section{Numerical results}
\label{section3}
In this section, we investigate the Hamiltonian (\ref{Hamiltonian}) 
by means of the exact-diagonalization method.
The exact-diagonalization
method has been utilized successfully in the course of the studies
of the $S=1/2$ random spin systems \cite{Haas93,Runge94,Runge92}.

\subsection{Suppression of the string correlation by the randomness}
\label{section3_1}

In Fig. \ref{string_0}, we plotted the string correlation 
\cite{den89,Tasaki91}
${\cal O}^{z}_{\rm string}(L/2)$ against the randomness $\Delta$
at the Heisenberg point $\beta=0$.
\begin{figure}[htbp]
\begin{center}\leavevmode
\epsfxsize=10cm
\epsfbox{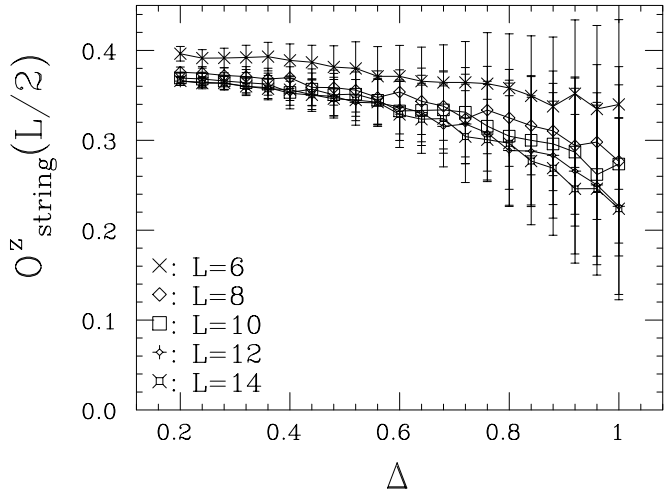}
\end{center}
\caption{
The string correlation ${\cal O}_{\rm string}^z(L/2)$ 
({\protect \ref{string_correlation}}) is plotted
for various randomness $\Delta$ at the Heisenberg point $\beta=0$.
The string correlation which characterizes the Haldane phase 
becomes suppressed as the randomness is strengthened.
}
\label{string_0}
\end{figure}
The string correlation is defined as
\begin{equation}
{\cal O}^z_{\rm string}(j-i)=
\left[
\langle S^z_i {\rm e}^{{\rm i}\pi\sum_{k=i}^{j-1} S^z_k} S^z_j \rangle
\right]_{\rm av}
 \ .
\label{string_correlation}
\end{equation}
The random average $[ \  ]_{\rm av}$ is taken over ninety samples
throughout the present study.
The string correlation was invented in order to detect characteristics
of the Haldane state.
The long-range limit was estimated numerically,
${\cal O}^\alpha_{\rm string}(r \to \infty) \to 0.38$,
at $\beta=0$ \cite{Girvin89}.
(It is noteworthy that the usual correlations,
such as the N{\'e}el correlation, are short-range.)
In Fig. \ref{string_0},
we see that the string order dominates in the weak-random region,
while it becomes suppressed as the randomness is strengthened.

In order to observe whether the string correlation is long-range or not,
we evaluate the Binder parameter \cite{Binder81} for the string correlation,
\begin{equation}
U(L)=1-\frac{[\langle O^4 \rangle]_{\rm av}}
            {3[\langle O^2 \rangle]_{\rm av}},
\label{Binder_parameter}
\end{equation}
where $O=\sum_{i=1}^{L} {\rm e}^{{\rm i}\pi\sum_{k=1}^{i-1}S^z_k} S^z_{i}$.
The Binder parameter is enhanced (suppressed)
with increasing the system size,
if the corresponding order is long-range (short-range).
It is plotted in Fig. \ref{Binder_0},
where the parameter varies in the same 
range as that in Fig. \ref{string_0}.
\begin{figure}[htbp]
\begin{center}\leavevmode
\epsfxsize=10cm
\epsfbox{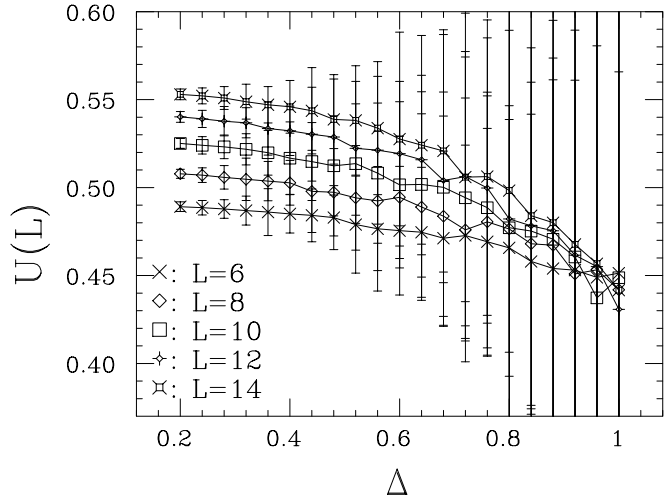}
\end{center}
\caption{
The Binder parameter $U(L)$ ({\protect \ref{Binder_parameter}})
is plotted
for the same parameter range as that shown in Fig. {\protect \ref{string_0}}.
The plot shows that the string order develops over the
whole region, and at the extreme randomness $\Delta=1$, it is disturbed.
}
\label{Binder_0}
\end{figure}
We see that the string order develops in the whole range of the randomness.
The result contradicts the picture \cite{Hyman97}
reviewed in the previous section.

As the biquadratic term is turned on ($\beta>0$),
the Haldane phase becomes suppressed.
(Note the phase diagram for the pure case reviewed 
in the section \ref{section1}.)
In Fig. \ref{Binder_t2}, where we plotted the Binder parameter at $\beta=0.2$.
\begin{figure}[htbp]
\begin{center}\leavevmode
\epsfxsize=10cm
\epsfbox{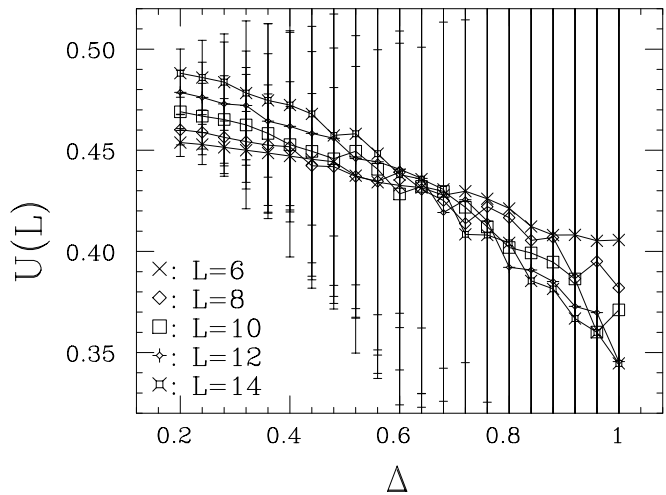}
\end{center}
\caption{
The Binder parameter $U(L)$ is plotted for various randomness
$\Delta$ at $\beta=0.2$.
We see an intersection point $\Delta\sim0.6$ 
which suggests a transition point.
}
\label{Binder_t2}
\end{figure}
It is implied that there exists a transition point $\beta\approx0.6$;
in one side $\beta<0.6$ the string order develops, while
in the other side $0.6<\beta$ it is disturbed.
Some might wonder that the result of $\beta=0.2$
in Fig. \ref{Binder_t2} is inconsistent with
the result of $\beta=0$ in Fig. \ref{Binder_0}, and the former supports
the picture 
of the previous theory \cite{Hyman97}.
In the following, however, we show an evidence
that the transition point (intersection point) is extrapolated
to $\Delta \approx 1$ in the thermodynamic limit $L\to\infty$.

\subsection{Analysis of the transition point}
\label{section3_2}

According to the theory \cite{Hyman97} reviewed in the section \ref{section2},
the transition should be the second order.
Namely, the Binder parameter (\ref{Binder_parameter}) obeys
the finite-size scaling hypothesis,
\begin{equation}
U(L)=L^x \tilde{U}
\left(
(\Delta-\Delta_{\rm c})L^{1/\nu}
\right)\ .
\label{scaling_theory}
\end{equation}
The scaling dimension $x$ is zero, because the Binder parameter
is dimensionless.
Therefore, the data $(\Delta-\Delta_{\rm c})L^{1/\nu})$-$U(L)$,
the so-called scaling data, must align along a curve which is
independent on the system size $L$.
We determine the scaling parameters $\Delta_{\rm c}$ and $\nu$
in eq. (\ref{scaling_theory}),
so that the data converge to such an universal
curve; see the appendix for details.

In Fig. \ref{scaling_plot_t2},
we show the scaling plots for $\beta=0.2$.
\begin{figure}[htbp]
\begin{center}\leavevmode
\epsfxsize=10cm
\epsfbox{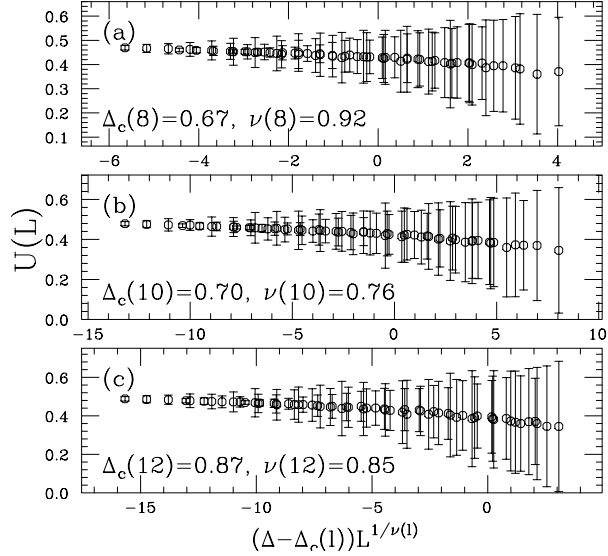}
\end{center}
\caption{
Scaling plots for $\beta=0.2$.
In order to observe the systematic corrections to the
finite size scaling,
we managed to analyse different three sets of the system sizes;
(a) $L=6,8,10$, (b) $L=8,10,12$ and (c) $L=10,12,14$, respectively.
}
\label{scaling_plot_t2}
\end{figure}
In order to see the systematic corrections to the finite size scaling,
the scaling analyses are managed for different three sets of the system sizes;
(a) $L=6,8,10$, (b) $L=8,10,12$ and (c) $L=10,12,14$, respectively.
We define the approximative transition point 
$\Delta(l)$ 
and the exponent
$\nu(l)$, which are
determined by the scaling plot $L=l-2,l$ and $l+2$.
We found that
the scaling parameters result in the following,
\begin{eqnarray}
\Delta_{\rm c}(8)=0.67,\  & \Delta_{\rm c}(10)=0.70,\  & 
\Delta_{\rm c}(12)=0.87,   \nonumber \\ 
\nu(8)=0.92,\ & \nu(10)=0.76,\ & \nu(12)=0.84.
\end{eqnarray}
The transition point converges to $\Delta\sim1$ for large system sizes,
and the exponent approaches $\nu\sim0.9$.
It is suggested that
the transition point is identical with that at the Heisenberg point
$\beta=0$, see Fig. \ref{Binder_0}.
% We conjecture that the transition point might locate at the
% extremely-broad-bond distribution $\Delta=1$.
% Note that the data for $\beta=0$ shown in Fig. \ref{Binder_0}
% does not indicate any finite-randomness transition.

We show the scaling plots for $\beta=0.3$ as well,
see Fig. \ref{scaling_plot_t3}.
\begin{figure}[htbp]
\begin{center}\leavevmode
\epsfxsize=10cm
\epsfbox{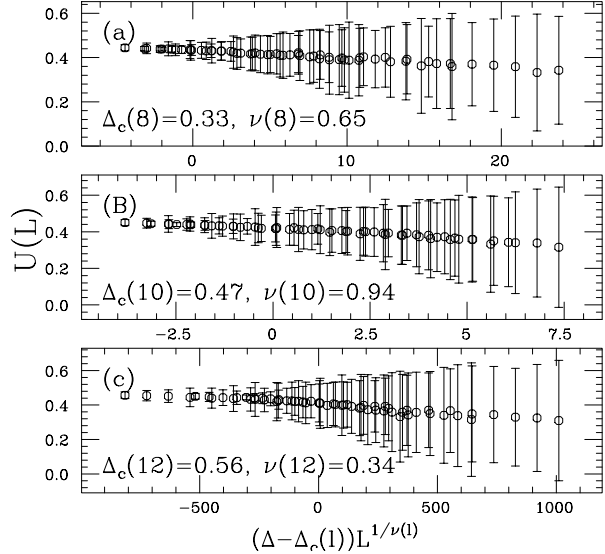}
\end{center}
\caption{
Scaling plots for $\beta=0.3$.
The analysis is the same as that shown in Figs. 
{\protect \ref{scaling_plot_t2}}.
}
\label{scaling_plot_t3}
\end{figure}
The plots yield the following estimates:
\begin{eqnarray}
\Delta_{\rm c}(8)=0.33,\  & \Delta_{\rm c}(10)=0.47,\  & 
\Delta_{\rm c}(12)=0.56,   \nonumber \\ 
\nu(8)=0.65,\ & \nu(10)=0.94,\ & \nu(12)=0.34.
\end{eqnarray}
We observe that the transition point approaches $\Delta \to 1$,
while the exponent rather scatters.
The scaling analyses beyond $\beta\sim0.3$ are suffered by strong finite-size
corrections which also exist for the uniform system 
(refer to the section \ref{section1}).
In consequence of the previous and the present
subsections, we observe that a phase transition might occur 
at $\Delta=1$.
The universality class is rather ambiguous.
It is suggested, at least, that the previous prediction 
$\nu\approx2.3$ \cite{Hyman97}
is not accordant with our numerical analysis.

In the next subsection,
we show an evidence that the transition is the first order.
Even for such case, the above finite-size-scaling
analysis is still meaningful in the sense that it yields
the estimate of the intersection point of the Binder parameters.
The intersection point is expected to converge to the transition point
even for the first-order transition.
The exponent, on the contrary, is not universal, and possibly would 
not be well-defined for very large system sizes.

% \subsection{Estimation of the transition point}
% If the transition is the first order, the analyses 
% in the previous subsection,
% where the scaling relation (\ref{scaling_theory}) is assumed,
% does not make sense.
% Hence, here, we demonstrate an alternative analysis 
% of the Binder-parameter curves such
% as those shown in Fig. \ref{Binder_t2}.
% We define the approximate transition point
% which is the intersection point of the curves of $L=l$ and $l+2$,
% The intersection point is determined
% so that the curves of the ????????????????? 

% \subsection{Barber-Batchelor region $\beta$}
%
% As the biquadratic interaction is applied furthermore,
% correction to scaling analysis.
% This difficulty also appears in the analyses for the pure system.
% The scaling plot for $\beta=$ is shown in Fig.
% The finite size correction is considerably large both for $\Delta_c$
% and $\nu$.
% The trend of $\nu$ is rather contrastive to the above estimate;
% It vanishes as the system size is enlarged.
% This tendency might be understood as follows.
% Note that the Barber-Batchelor phase is unstable;
% namely, the gapful phase becomes immediately distracted by an infinitesimal
% randomness.
% This critical phenomena is something like the one with $\nu=+0$.
% We thus suspect this criticality affect that of the Haldane-random-singlet 
% phase transition to some extent.

\subsection{Correlation length and the spin-stiffness constant}
\label{section3_3}

In order to confirm that a first-order transition takes place at
$\Delta=1$, we show the result of the correlation length.
We found that the correlation length is remained unchanged over the
whole region $0<\Delta<1$; namely, the correlation length does not
diverge as in eq. (\ref{correlation_exponent}).

In Fig. \ref{neel_0},
the averaged N{\'e}el correlation  
is plotted for various randomness at the Heisenberg point $\beta=0$.
\begin{figure}[htbp]
\begin{center}\leavevmode
\epsfxsize=10cm
\epsfbox{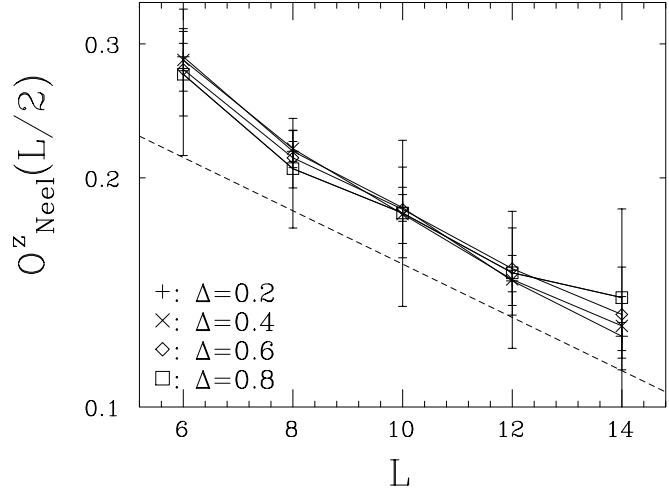}
\end{center}
\caption{
Logarithm of the
N{\'e}el correlation ${\cal O}^z_{\rm Neel} (L/2)$
is plotted
against the system size $L$
for various randomness at the Heisenberg point $\beta=0$.
The decay rate ($\sim {\rm e}^{-r/6.2}$) at $\Delta=0$ 
{\protect \cite{Golinelli94}} is shown by the dashed line.
The plot shows that the correlation decays
exponentially in fact, and the correlation length is not changed very
much with increasing the randomness.
}
\label{neel_0}
\end{figure}
The averaged N{\'e}el correlation is given by
\begin{equation}
{\cal O}^z_{\rm Neel}(j-i)=\left[ \langle 
(-1)^{j-i}S^z_i S^z_j \rangle \right]_{\rm av}.
\end{equation}
The decay rate $\sim {\rm e}^{-(j-i)/6.2}$ estimated previously
\cite{Golinelli94} at $\Delta=0$ is shown as well.
The magnetic correlation length is hardly changed with the randomness.
Moreover,
we observe that the correlation itself is not changed very much.
The result indicates that the Haldane phase with finite correlation length
$\xi\sim6.2$
continues up to the extreme point $\Delta=1$.

In Fig. \ref{stiffness_t0}, the spin stiffness constant
is plotted.
\begin{figure}[htbp]
\begin{center}\leavevmode
\epsfxsize=10cm
\epsfbox{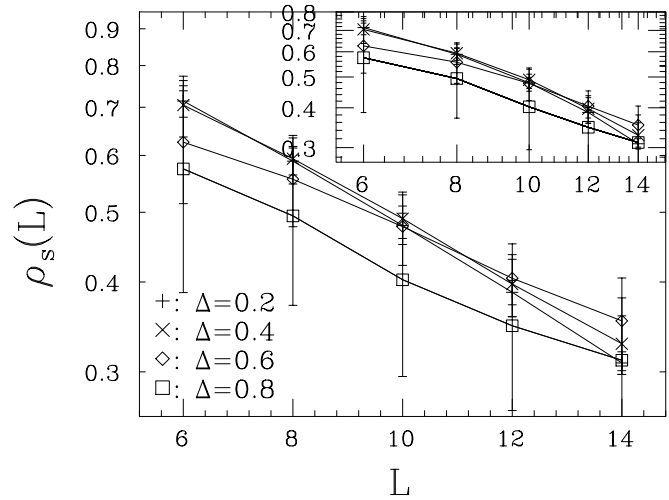}
\end{center}
\caption{
Logarithm of the
spin-stiffness constant $\rho_{\rm s}(L)$
is plotted
against (a) $L$ and (b) $\log L$, respectively,
for various randomness at $\beta=0$.
}
\label{stiffness_t0}
\end{figure}
The stiffness constant \cite{Fisher73} is defined by
\begin{equation}
\rho_{\rm s}=L\frac{\partial^2 E_{\rm g}(L)} {\partial \theta^2},
\end{equation}
where $\theta$ denotes the boundary twist,
\begin{equation}
{\bf S}_L \cdot {\bf S}_1 \to 
\frac12(S^+_L S^-_1 {\rm e}^{{\rm i}\theta}
       +S^-_L S^+_1 {\rm e}^{-{\rm i}\theta})
+S^z_L S^z_1 .
\end{equation}
The stiffness constant is generally employed for investigating
the random system:
If the randomness destroys long-range coherence,
the stiffness should vanish.
In the Haldane phase, the stiffness vanishes exponentially 
with enlarging the  system size,
because the ground state has short-range magnetic correlation.
We observe in Fig. \ref{stiffness_t0} that the stiffness 
actually decays exponentially
over the while randomness with the localization length unchanged.
In consequence, we conclude that in the Haldane phase, the ground state
property is kept unchanged against the randomness,
and it transits at the point 
$\Delta=1$ abruptly.
We will discuss the transition scenario
in the next section.

% \subsection{Spin-stiffness}
% Here, we present the spin-stiffness constant
% \begin{equation}
% \frac{\partial}{\partial}.
% \end{equation}
% It is generally employed for random systems as that in the present case.
% It should vanish for such systems as localized phases induced by
% impurities.
% 
% In Fig. the stiffness constant is plotted against the inverse of the 
% system sizes for several conditions.
% The constant at $\beta$ seems to vanishes through the thermodynamic limit
% while that at $\beta$ remains finite.
% The result supports the phase diagram estimated in the previous subsection.
% Note that the former condition is located in the Haldane phase,
% the stiffness should be vanishing in this phase.
% On the other hand, the stiffness in the random singlet phase 
% remains finite.

\section{Summary and discussions}
\label{section4}

We summarize and discuss our numerical results.
The $S=1$ Heisenberg chain (\ref{Hamiltonian})
with the bond randomness $\Delta$
is studied numerically.
The Haldane phase $(\beta\approx0)$ survives over the whole random region
$0<\Delta<1$:
The intersection point of the curves of the Binder parameter
converges to the point $\Delta \approx 1$ in the thermodynamic limit.
In fact, the result is consistent with the previous theory
\cite{Hyman96,Yang96},
which suggests that the Haldane phase
would be stable against randomnesses considerably,
because the Haldane state has such a hidden correlation
 as the string correlation
(\ref{string_correlation}).
(Hence, the theory \cite{Hyman97} reviewed 
in the section \ref{section2} somehow contradicts the above.)
The present system is, thus, quite contrastive to the 
Heisenberg ladder with the non-magnetic-impurities,
whose spin-liquid ground state is unstable against infinitesimal
doping \cite{Fukuyama96,Martins96,Mikeska97}.

We observed that the correlation length is hardly changed with
the randomness increased.
Hence, we conclude that the correlation length does not
diverge in the vicinity of
 $\Delta=1$, and thus the phase transition is the first order.
(Note that according to the theory \cite{Hyman97}, a second-order transition
separates the Haldane phase and the random-singlet phase at
a critical randomness 
$\Delta_{\rm c}$.)
At the transition point $\Delta=1$, namely, the extremely-%
broad-bond distribution, infinitesimally weak bonds appear.
It is thereby expected that
at this very point the random-singlet phase is realized \cite{Monthus97}.

Finally, we mention what phase extends beyond the point $\Delta>1$.
In the region, ferromagnetic bonds appear as well.
This sort of randomness is considered for the $S=1/2$ Heisenberg chain,
see the article \cite{Nguyen96} for a review.
We suspect that in this region, 
the ground state is similar to that realized for the 
the $S=1/2$ chain:
It is found that the low-temperature properties resemble those
of the classical counterpart \cite{Tonegawa75}.
This similarity is quite convincing, because the ferromagnetism
is not affected very much by quantum fluctuation.
It is natural that at least the magnitude of the spin 
does not alter the ground-state property.
% (In actual, this is why we did not investigated the region $\Delta>1$.
% Namely, ferromagnetism violate the uniqueness of the ground state.
% Thereby the numerical analyses face troubles.)
The Haldane phase $\Delta<1$, on the contrary,
owes the stabilization to the zero-point quantum fluctuation.
In consequence, we see that at $\Delta=1$ a drastic transition
occurs from the Haldane phase to the above rather classical phase.

\section*{Acknowledgement}
Our computer programs
are partly based on the subroutine package "TITPACK Ver. 2"
coded by Professor H. Nishimori.
The numerical calculations were performed on the super-computer HITAC 
S3800/480 of the computer centre, University of	Tokyo, and on the 
work-station HP Apollo 9000/735 of the Suzuki group, Department of 
Physics University of Tokyo.

\appendix

\section{Details of the present scaling analyses}
\label{appendix}

We explain the details of our finite-size-scaling analyses,
which we managed in the section \ref{section3_2}
in order to estimate the transition point $\Delta_{\rm c}$ and
the exponent $\nu$.
We adjusted these scaling parameters
 so that the scaled data, such as those shown in
Figs. \ref{scaling_plot_t2} and \ref{scaling_plot_t3},
form a curve irrespective of the system sizes.
In order to see qualitatively to what extent these data align,
we employ the ``local lineality function'' $S(\Delta_{\rm c},\nu)$
defined by Kawashima and Ito \cite{Kawashima93}:
Suppose
a set of the data points $\{ (x_i,y_i) \}$ with the errorbar
$\{ d_i(=\delta y_i) \}$, which
we number so that $x_i<x_{i+1}$ may
hold for $i=1,2,\cdots,n-1$.
For this data set, the local-lineality function is defined as
\begin{equation}
S=\sum_{i=2}^{n-1} 
w(x_i,y_i,d_i|x_{i-1},y_{i-1},d_{i-1},x_{i+1},y_{i+1},d_{i+1}).
\label{linear}
\end{equation}
The quantity $w(x_j,y_j,d_j|x_i,y_i,d_i,x_k,y_k,d_k)$
is given by
\begin{equation}
w=
\left(
\frac
{y_j-\bar{y}}
{\Delta}
\right)^2 \ ,
\end{equation}
where
\begin{equation}
\bar{y}=\frac{(x_k-x_j)y_i-(x_i-x_j)y_k}
{x_k-x_i}
\end{equation}
and
\begin{equation}
\Delta^2=d^2_j+
\left(
\frac{x_k-x_j}{x_k-x_i} d_i
\right)^2
+
\left(
\frac{x_i-x_j}{x_k-x_i}d_k
\right)^2.
\end{equation}
In other words, the numerator $y_j-\bar{y}$ 
denotes the deviation of the point 
$(x_j,y_j)$
from the line passing
two points $(x_i,y_i)$ and $(x_k,y_k)$, and
the denominator $\Delta$ stands for the statistical error
of $(y_i-\bar{y})$.
And so, $w=((y_i-\bar{y})/\Delta)^2$ shows a degree to what extent
these three points align.

As the number of the data points $n$ is increased,
the statistical error of $S$ would be reduced.
The corrections to finite-size scaling
might increase instead.
In the present analyses, we used twenty data in the vicinity of the
transition point $\Delta_{\rm c}$.
An example of the plot $S$ is shown in Fig. \ref{lineality}.
\begin{figure}[htbp]
\begin{center}\leavevmode
\epsfxsize=10cm
\epsfbox{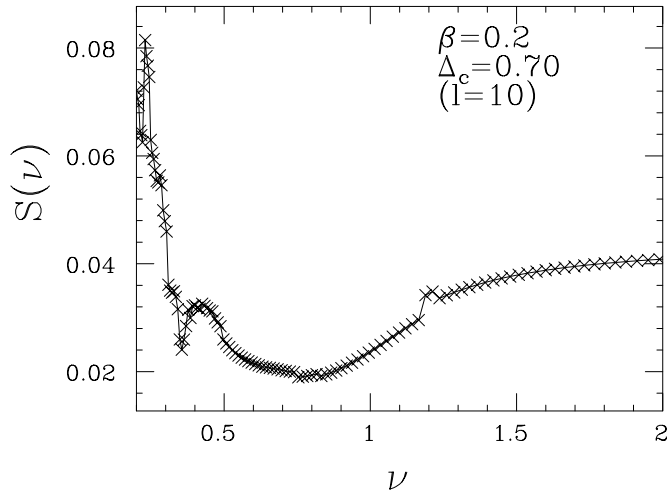}
\end{center}
\caption{
The local linearity function $S$ ({\protect \ref{linear}})
for the scaling data in Fig. {\protect \ref{scaling_plot_t2}} (b)
is plotted for various $\nu(l=10)$
with $\Delta_{\rm c}=0.70$ fixed.
The location of the minimum yields the estimate of the exponent $\nu$.
}
\label{lineality}
\end{figure}
We observe the minimum at $\nu\approx 0.76$,
which yields the estimate of the exponent.

\end{large}
\end{document}